\documentclass[aps,prl,twocolumn,showpacs,superscriptaddress,groupedaddress]{revtex4}  
\pdfoutput=1
\usepackage{amssymb}   
\usepackage{amsmath}
\usepackage{graphicx}

\def\beq{\begin{equation}}
\def\eeq{\end{equation}}
\def\bsp#1\esp{\begin{split}#1\end{split}}

\begin{document}

\begin{flushleft} 
\mbox{CERN-PH-TH/2015-055, CP3-15-07}
\end{flushleft}

\title{Higgs boson gluon-fusion production in N$^3$LO QCD}

\author{Charalampos Anastasiou}
\affiliation{Institute for Theoretical Physics, ETH Z\"urich,
  8093 Z\"urich, Switzerland}

\author{Claude Duhr}
\altaffiliation{On leave from the ``Fonds National de la Recherche Scientifique'' (FNRS), Belgium.}
\affiliation{CERN Theory Division, 1211 Geneva 23, Switzerland}
\affiliation{Center for Cosmology, Particle Physics and Phenomenology (CP3),
Universit\'{e} Catholique de Louvain, 1348 Louvain-La-Neuve, Belgium} 

\author{Falko Dulat}
\affiliation{Institute for Theoretical Physics, ETH Z\"urich,
  8093 Z\"urich, Switzerland}

\author{Franz Herzog}
\affiliation{Nikhef, Science Park 105, 1098 XG Amsterdam, The Netherlands}

\author{Bernhard Mistlberger}
\affiliation{Institute for Theoretical Physics, ETH Z\"urich,
  8093 Z\"urich, Switzerland}

\date{\today}

\begin{abstract}
We present the cross-section for  the production of a Higgs boson at hadron-colliders  
at next-to-next-to-next-to-leading order (N$^3$LO) in perturbative
QCD.  The calculation is based on a method to perform a series
expansion of the partonic cross-section around the threshold limit to
an arbitrary order. We perform this expansion to sufficiently high order to obtain the value of the hadronic cross at N$^3$LO in the large top-mass limit. 
For renormalisation and factorisation
scales equal to half the Higgs mass, the N$^3$LO corrections are of the order
of $+2.2\%$.  The total scale variation at N$^3$LO is $3 \%$, reducing the
uncertainty due to missing higher order QCD corrections by a factor of
three.  
\end{abstract}

\maketitle

The success of the Large Hadron Collider (LHC) experiments in the
exploration and interpretation of phenomena  at the TeV scale is due, on the one hand, to amazing
experimental and technological advances and, on the other hand, extraordinary
progress in perturbative QCD. 
In particular, the discovery of the Higgs boson~\cite{HiggsDiscovery}
 by the LHC experiments has initiated an era of precision studies of the
 properties of the Higgs boson, where precise theory predictions for
 Higgs observables  play an indispensable role. 

The inclusive gluon-fusion cross-section is a prototypical example of
a theoretical input for the interpretation of the experimental
observations. 
It enters not only into the extraction of the Higgs-boson couplings
from the measurements, but it could also play an important role in
identifying deviations from the Standard Model predictions in Higgs physics.
  Unfortunately,
the theory predictions for the inclusive cross-section 
are plagued by significant theoretical uncertainties.
Scale variations at NNLO indicate that 
 missing higher order effects are of the order of $\pm 9 \%$ at LHC 
energies~\cite{ihixs8tev,ihixs}, and the size of this uncertainty is comparable to the
experimental uncertainty from the LHC Run 1~\cite{Khachatryan:2014jba,Aad:2014eha}. 
Hence, 
with a few more years of data taking the theoretical uncertainty will
be dominant, demanding for an update of the current theoretical predictions.

In this context, a vigorous effort has recently been made to compute the
inclusive gluon-fusion cross-section at next-to-next-to-next-to-leading order (N$^3$LO) in perturbative QCD.  
The cross-section at N$^3$LO receives contributions from many
different building blocks, all of which have been computed over the
last years, at least partially. The three-loop corrections to Higgs
production in gluon fusion have been obtained in
ref.~\cite{formfactor}, and the corrections from the emission of an
additional parton at one or two loops were computed in
ref.~\cite{Anastasiou:2013mca,Kilgore:2013gba,Gehrmann:2011aa,Duhr:2013msa,Duhr:2014nda}. In
order to obtain a finite result, appropriate ultra-violet and infrared
counterterms need to be included~\cite{UV,IR,NNLOXsec}. While all of
these contributions had been computed in full generality,
contributions from the emission of two partons at one loop and three
partons at tree-level had only been computed in an approximate
manner. In particular, for these contributions the first two terms in
the expansion around threshold could be
obtained~\cite{triplereal,Zhu:2014fma,Anastasiou:2014vaa,Li:2014bfa}, 
confirming previous results for logarithmically enhanced terms in the
cross section~\cite{Moch:2005ky} 
and 
resulting in the complete computation of the inclusive gluon-fusion cross-section at N$^3$LO in the soft-virtual~\cite{Anastasiou:2014vaa,Li:2014bfa,Li:2014afw} and next-to-soft approximations~\cite{Anastasiou:2014lda}. Owing to the universality of soft emissions, the previous results have sparked various new results for QCD processes at N$^3$LO in the soft-virtual approximation~\cite{otherSV}.

Despite this progress, the soft-virtual and next-to-soft
approximations are insufficient to make reliable predictions for the
cross-section, owing to a slow convergence of the threshold
expansion~\cite{Anastasiou:2014lda}. In this Letter we close this gap,
and we present the gluon-fusion Higgs production cross-section at
N$^3$LO in perturbative QCD. We emphasise that this is the first ever
complete computation of a cross-section at N$^3$LO at a hadron
collider. 

We will describe the main result of our computation in this Letter, 
while  a detailed account of the mathematical and computational methods will be presented elsewhere. Here it suffices to say that we work in the framework of reverse-unitarity~\cite{nnlobabis,reverse-unitarity}, and we perform a complete reduction of the cross-section to master integrals, without
any approximations.
For the double and triple-emission contributions at N$^3$LO, we can derive differential equations satisfied by
the master integrals~\cite{nnlobabis,reverse-unitarity,DiffEqs}, which we solve as generalised power series around the threshold limit.  
In this way, we obtain at least 37 terms in the threshold expansion of each master integral.
An important part of our computation has been the evaluation of the
boundary conditions which are needed for solving the differential
equations for the master integrals. 
Many of the boundary conditions required in this project had already been derived 
in the context of the soft-virtual and next-to-soft results~\cite{triplereal,Zhu:2014fma,Anastasiou:2014vaa,Li:2014bfa,Anastasiou:2014lda}. Using similar techniques, 
we have computed the remaining few unknown boundary conditions for master integrals which start to be relevant only at a high order
in the threshold expansion.

Having at our disposal the complete set of master integrals as expansions around the threshold limit, we can easily obtain the cross-sections at N$^3$LO for 
all partonic channels contributing to Higgs production via gluon fusion. The partonic cross-sections are related to the hadronic cross-section at the LHC through the integral
\begin{equation}
\label{eq:sigmahadronic}
\sigma = \sum_{i,j} \int dx_1 dx_2 f_i(x_1,\mu_f) f_j(x_2,\mu_f)
\hat{\sigma}_{ij}(z,\mu_r, \mu_f)\,, 
\end{equation}
where the summation indices $i,j$ run over the parton flavors in the
proton, $f_i$ are parton densities and $\hat{\sigma}_{ij}$ are partonic
cross-sections. Furthermore, we define $z=\frac{m_H^2}{s}$, where $m_H$ is the mass of the Higgs boson
and $\sqrt{s}$ is the partonic center-of-mass energy, related to the hadronic center-of-mass energy $\sqrt{S}$ through $s = x_1\,x_2\,S$. 
The renormalisation and factorisation scales are denoted by $\mu_r$ and $\mu_f$.  
We work in an effective theory approach where the top-quark is
integrated out.  The  effective Lagrangian describing the interaction
of the Higgs boson and the gluons is, 
\begin{equation}
\label{eq:effectivelagrangian}
{\cal L}_{\textrm{eff}} = -\frac{C}{4} H \, G_{\mu \nu }^a G^{a \mu \nu}\,,
\end{equation}
where $H$ is the Higgs field, $G^a_{\mu \nu}$ is the gluon field strength tensor and $C$ the Wilson
coefficient, known up to N$^4$LO~\cite{Wilson}.  We expand the
partonic cross-sections into a perturbative series in the strong coupling constant evaluated at the scale $\mu_r$,
\begin{equation}
\label{eq:alphaexpansion}
\hat{\sigma}_{ij} =  \hat{\sigma}_0\,\left[\delta_{ig}\,\delta_{jg}\,\delta(1-z) +\sum_{\ell=1}^\infty\left(\frac{\alpha_s(\mu_r)}{\pi}\right)^\ell \hat{\sigma}_{ij}^{(\ell)}\right]\,.  
\end{equation}
 In this expression $\hat{\sigma}_0$ denotes the leading order cross-section, and the terms through NNLO in the above expansion have been computed
in~\cite{nlo,nnlobabis,nnlo}.  The main result of this Letter is the result for the N$^3$LO coefficient, corresponding to $\ell=3$ in eq.~\eqref{eq:alphaexpansion}, for all possible parton flavours in the initial state. We cast the N$^3$LO coefficients in the form
\begin{equation}
\label{zbarexpansion}
\hat{\sigma}_{ij}^{(3)} = \lim_{N \to \infty}  \hat{\sigma}_{ij}^{(3,N)}\,, 
\end{equation} 
where we introduce the truncated threshold expansions defined by
\begin{equation}
\label{eq:partialsum}
\hat{\sigma}_{ij}^{(3,N)}=\delta_{ig}\,\delta_{jg}\,\hat{\sigma}_{\textrm{SV}}^{(3)} + \sum_{n=0}^{N} c_{ij}^{(n)} \, (1-z)^{n}\,. 
\end{equation}
Here, $\hat{\sigma}_{\textrm{SV}}^{(3)}$ denotes the soft-virtual cross-section at N$^3$LO of ref.~\cite{Anastasiou:2014vaa,Li:2014bfa,Li:2014afw} and $N=0$ is the next-to-soft approximation of ref.~\cite{Anastasiou:2014lda}. Using our method for the threshold expansion of the master integrals, we were able to determine the  $c_{ij}^{(n)}$ analytically up to at least $n = 30$. Note that at any given order in the expansion these coefficients are polynomials in $\log(1-z)$.
\begin{figure}[!t]
\begin{center}
\includegraphics[width=0.48\textwidth]{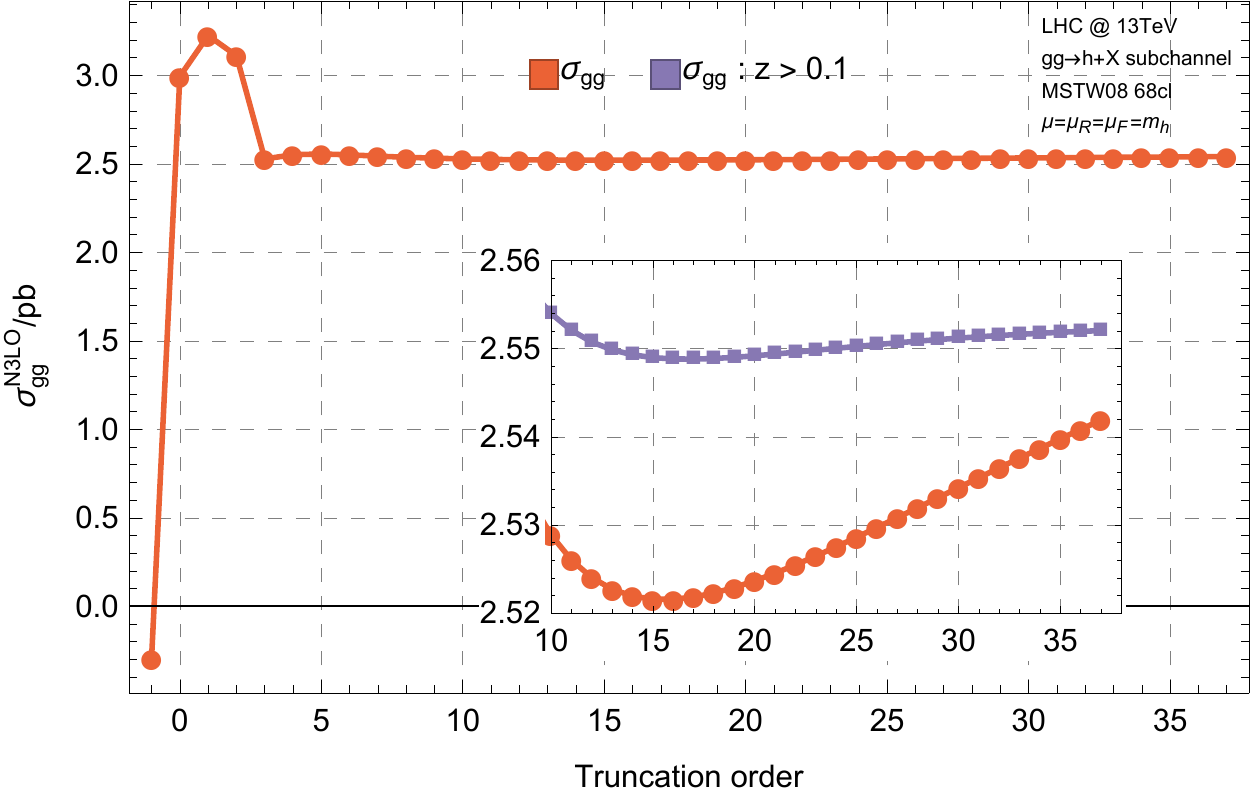}
\end{center}
\caption{
\label{fig:zbarexpansionnumerics}
The N$^3$LO correction from the $gg$ channel to the hadronic cross-section as a
  function of the truncation order $N$ in the threshold expansion for the scale choice $\mu=m_H$.}
\end{figure}
While this approach does not cast the partonic cross-sections in a
closed analytic form, we argue that it yields the complete result for the value of the hadronic
cross-section. 
In Fig.~\ref{fig:zbarexpansionnumerics} we show the contribution of
the partonic cross-section coefficients N$^3$LO to
the hadronic cross-section for
a proton-proton collider with $13\, {\rm TeV}$ center-of-mass energy as a
function of the truncation order $N$.  We 
use NNLO MSTW2008~\cite{mstw} parton densities and a value for the
strong coupling at the mass of the $Z$-boson of $\alpha_s(m_Z)=0.117$ as initial value for the evolution, and we set the factorisation scale to $\mu_f=m_H$.  
We observe that the threshold expansion stabilises starting from $N=4$,  leaving a negligible truncation uncertainty for the hadronic cross-section thereafter.   
We note, though, that we observe a very small, but systematic,
increase of the expansion in the range $N\in[15,37]$, as illustrated
in Fig.~\ref{fig:zbarexpansionnumerics}. We have observed that a
similar behaviour is observed for the threshold expansion at NNLO. 
The systematic increase originates from values of the partonic
cross-section at very small $z$.  Indeed, this increase appears only
in the contributions to the hadronic cross-section integral for values
$z < 0.1$.  It is natural that the terms of the threshold expansion
computed here do not furnish a good approximation of the
hadronic integral in the small $z$ region  due to the divergent high
energy behaviour of the partonic cross sections
~\cite{Hautmann:2002tu}.  However, it is observed that this region is suppressed in the
total hadronic integral and for $z<0.1$ contributes less than $0.4\%$
of the total N$^3$LO correction. The same region at NLO and NNLO, where
analytic expressions valid for all regions are known, is similarly
supprerssed.  We therefore believe that the
uncertainty of our computation for the hadronic cross-section due to
the truncation of the threshold expansion is negligible (less than $0.2\%$). 

\begin{figure}[!t]
\begin{center}
\includegraphics[width=0.48\textwidth]{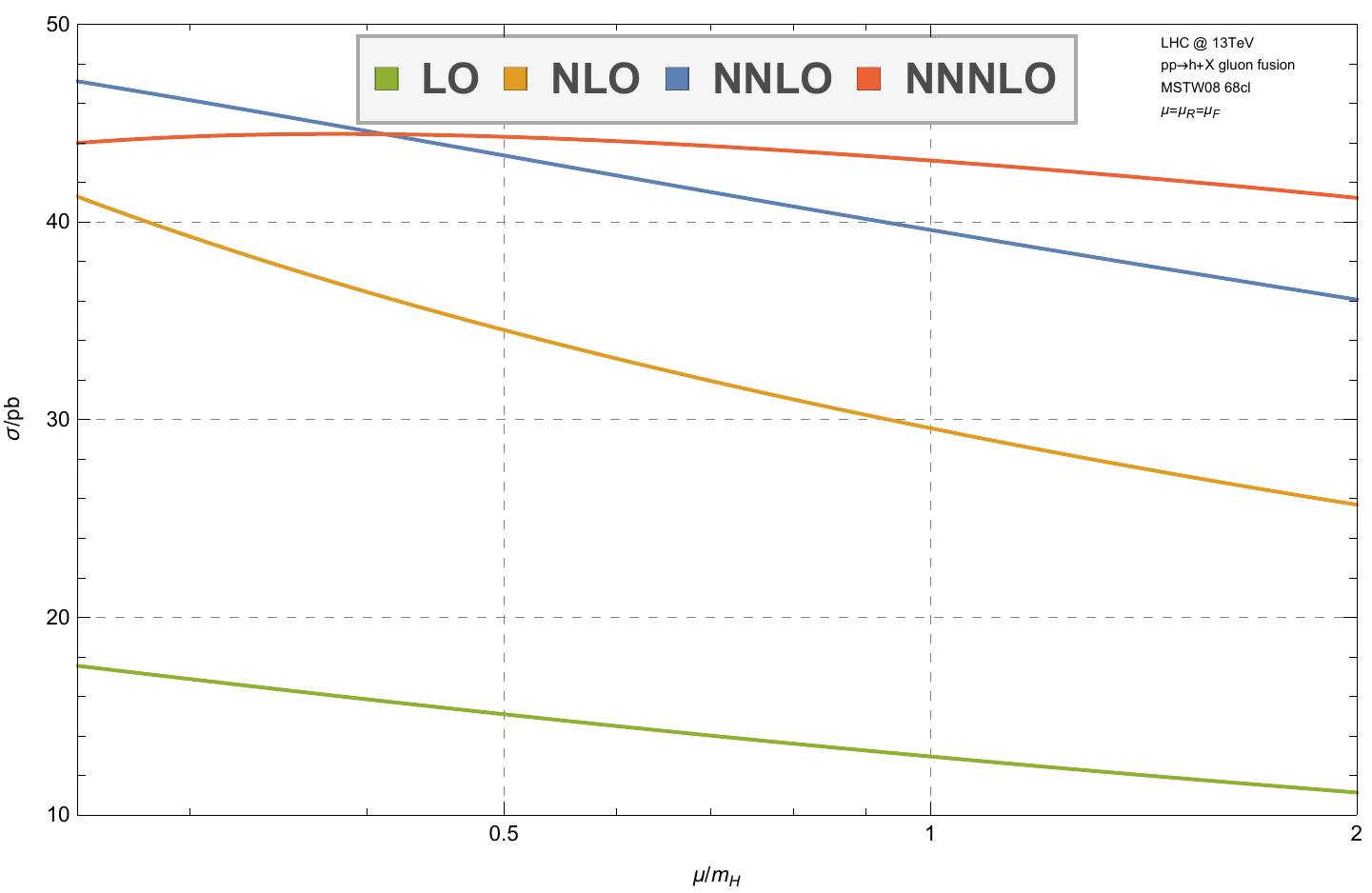}
\end{center}
\caption{
\label{fig:scalevariation}
Scale variation of the gluon fusion cross-section  at all perturbative
orders through N$^3$LO. 
}
\end{figure}
\begin{figure}[!t]
\begin{center}
\includegraphics[width=0.48\textwidth]{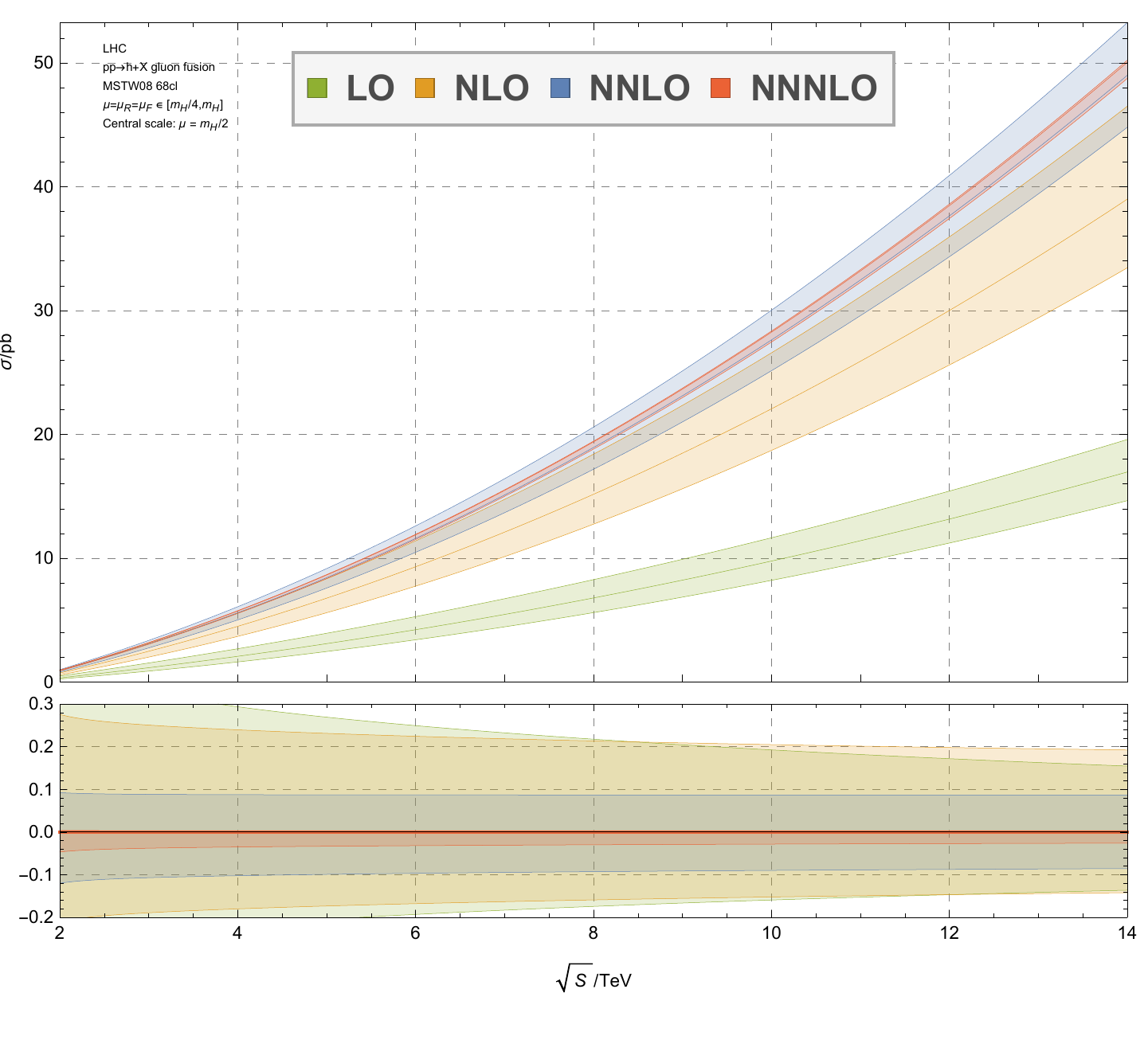}
\end{center}
\caption{
\label{fig:energyvariation}
The gluon fusion cross-section  at all perturbative
orders through N$^3$LO in the scale interval $[\frac{m_H}{4}, m_H]$ as a
function of the center-of-mass energy $\sqrt S$. 
}
\end{figure}

In Fig.~\ref{fig:scalevariation} we present the hadronic gluon-fusion
Higgs production cross-section at N$^3$LO as a function of a common
renormalisation and factorisation scale $\mu = \mu_r = \mu_f$.  
We observe a significant reduction of the
sensitivity of the cross-section to the scale $\mu$. Inside a range $\mu
\in \left[ \frac{m_H}{4}, m_H\right]$ the cross-section at N$^3$LO varies in the interval $\left[-2.7\%,+0.3\%\right]$ with respect to the cross-section value at the central scale
$\mu=\frac{m_H}{2}$. For comparison, we note that the corresponding scale variation at NNLO  is about
$\pm 9\%$~\cite{ihixs8tev,ihixs}. This improvement in the precision of the Higgs
cross-section is a major accomplishment due to our calculation and will
have a strong impact on future measurements of Higgs-boson properties.  
Furthermore, even though for the scale choice $\mu = \frac{m_H}{2}$ the 
N$^3$LO corrections change the cross-section by about $+2.2\%$,
this correction is captured by the scale variation estimate for the missing higher
order effects of the NNLO result at that scale. We illustrate this point in Fig.~\ref{fig:energyvariation},
where we present the hadronic cross-section as a function of the hadronic center-of-mass energy $\sqrt{S}$ at the scale 
$\mu=\frac{m_H}{2}$.
We observe that the N$^3$LO scale uncertainty band is included within the
NNLO band, indicating that the perturbative expansion of the hadronic cross-section is convergent.  
However, we note that for a larger scale choice, e.g., $\mu =
m_H$, the convergence of the perturbative series is slower than for $\mu=\frac{m_H}{2}$.

In table \ref{tab:xs} we quote the gluon fusion cross section in effective theory at N$^3$LO for different LHC energies. The perturbative uncertainty is determined by varying the common renormalisation and factorisation scale in the interval $\left[\frac{m_H}{4},m_H\right]$ around $\frac{m_H}{2}$ and in the interval $\left[\frac{m_H}{2},2 m_H\right]$ around $m_H$.

\begin{widetext}
\center
\begin{table}[!h]
\begin{center}
\begin{tabular}{c|c c c c c}
\hline\hline
$\sigma/$pb & 2 TeV & 7 TeV & 8 TeV & 13 TeV & 14 TeV \\
\hline
\\[-10pt]
$\mu=\frac{m_H}{2}$ & $0.99^{+0.43\%}_{-4.65\%}$ & $15.31^{+0.31\%}_{-3.08\%}$ & $19.47^{+0.32\%}_{-2.99\%}$ & $44.31^{+0.31\%}_{-2.64\%}$ & $49.87^{+0.32\%}_{-2.61\%}$ \\[5pt]
$\mu=m_H$ & $0.94^{+4.87\%}_{-7.35\%}$ & $14.84^{+3.18\%}_{-5.27\%}$ & $18.90^{+3.08\%}_{-5.02\%}$ & $43.14^{+2.71\%}_{-4.45\%}$ & $48.57^{+2.68\%}_{-4.24\%}$ \\[5pt]
\hline\hline
\end{tabular}
\end{center}
\caption{
\label{tab:xs}
The gluon fusion cross-section in picobarn in the effective theory for different collider energies in the interval $[\frac{m_H}{4},m_H]$ around $\mu=\frac{m_H}{2}$ and in the interval $[\frac{m_H}{2},2m_H]$ around $\mu=m_H$.
}
\end{table}
\end{widetext}
Given the substantial reduction of the scale uncertainty at N$^3$LO, the
question naturally arises whether other sources of theoretical
uncertainty may contribute at a similar level. In the remainder of this 
Letter we briefly comment on this issue, leaving a more detailed quantitative study
for future work.

First, we note that given the small
size of the N$^3$LO corrections
compared to NNLO, we expect that an estimate for the higher-order
corrections at N$^4$LO and beyond can be obtained from the scale variation uncertainty.  
Alternatively, partial N$^4$LO results can be obtained by means of factorisation
theorems for threshold resummation.  However, we expect that the insight from
resummation on the N$^4$LO soft
contributions is only qualitative given the importance of
next-to-soft, next-to-next-to-soft and purely virtual contributions observed at 
N$^3$LO, as seen in Fig.~\ref{fig:zbarexpansionnumerics}. 
 
Electroweak corrections to Higgs production have been calculated through two loops
in ref.~\cite{ewk}, and estimated at three loops in
ref.~\cite{ewk3loop}. They furnish a correction of less than $+5\%$
to the inclusive cross-section.  Thus, they  are not negligible 
at the level of accuracy indicated by the scale variation at N$^3$LO
and need to be combined with our result in the future. 
Mixed QCD-electroweak or
purely electroweak corrections of even higher order are expected to contribute at the
sub-percent level and should be negligible.  

Next, we have to comment on our assumption that the top-quark is
infinitely heavy and can be integrated out, see eq.~\eqref{eq:effectivelagrangian}.
Moreover, we assumed that all other quarks have a zero Yukawa coupling. 
Finite quark mass effects are important, but it is sufficient that
they are inlcuded through NLO or NNLO.  
Indeed, finite quark-mass effects have been computed
fully through NLO in QCD~\cite{nlo}, while subleading top-quark mass corrections 
have been computed at NNLO systematically as an expansion in  the
inverse top-quark mass~\cite{nnlotop}.  In these references it was
observed that through NLO finite quark mass effects amount to about $8
\% $ of the K-factor.  At NNLO, the known $\frac{1}{m_{\rm top}}$
corrections affect the cross-section at the $\sim 1 \%$ level.  
A potentially significant contribution at NNLO which has not yet been
computed in the literature originates from diagrams with both a top
and bottom quark Yukawa coupling. 
Assuming a similar perturbative pattern as for top-quark only
diagrams in the effective theory, 
eq.~\eqref{eq:effectivelagrangian}, higher-order effects 
could be of the order  of $2\%$. We thus conclude that the computation 
of the top-bottom interference through NNLO is highly desired in the near future. 

Finally, the computation of the hadronic cross-section relies crucially on  
the knowledge of the strong coupling
constant and the parton densities.  After our calculation, the uncertainty
coming from these quantities
has become dominant. Further progress in the determination of
parton densities must be anticipated in the next few years due to the 
inclusion of LHC data in the global fits and the impressive advances
in NNLO computations, improving the theoretical accuracy of many
standard candle processes.  

To conclude, we have presented in this Letter the computation
of the gluon-fusion Higgs production cross-section through N$^3$LO in perturbative 
QCD. While a thorough study of the impact of electroweak and quark
mass effects is left for future work,  we expect that the remaining
theoretical uncertainty on the inclusive Higgs production
cross-section is expected to be reduced to roughly half, 
which will bring important benefits in 
the study of the properties of the Higgs boson at the LHC Run 2. 
Besides its direct phenomenological impact, we believe that 
our result is also a major advance in our understanding of perturbative QCD,
as it opens the door to push the theoretical predictions for large classes
of inclusive processes to N$^3$LO accuracy, like Drell-Yan
production, associated Higgs production and Higgs production via bottom
fusion. Moreover, on the more technical side, our result constitutes the first independent
validation of the gluon splitting function at NNLO~\cite{IR}, because the latter is required to cancel all the infrared poles 
in the inclusive cross-section. In addition, we expect that the
techniques developed throughout this work are not restricted to inclusive cross-sections, but it should be possible to extend them to certain classes of differential distributions, like rapidity distributions for Drell-Yan and Higgs production, thereby paving the way to a new era of precision QCD.

{\bf Acknowledgements:}
We are grateful to Elisabetta Furlan, Thomas Gehrmann and A. Lazopoulos 
for our collaboration on the many aspects  of the Higgs
cross-section N$^3$LO project
which are not covered in this Letter. We thank A. Lazopoulos in
particular for an independent implementation of our results in {\tt
  ihixs} and numerical comparisons.  
Research supported by the Swiss National Science Foundation (SNF) under 
contract 200021-143781 and the European Commission through 
the ERC grants ``IterQCD'', ``HEPGAME'' and ``MathAm''.



\begin{thebibliography}{99}

 



\bibitem{HiggsDiscovery}
  G.~Aad {\it et al.}  [ATLAS Collaboration],
  Phys.\ Lett.\ B {\bf 716}, 1 (2012);
  S.~Chatrchyan {\it et al.}  [CMS Collaboration],
  Phys.\ Lett.\ B {\bf 716}, 30 (2012).
  


\bibitem{ihixs8tev} 
  C.~Anastasiou, S.~B\"uhler, F.~Herzog and A.~Lazopoulos,
  JHEP {\bf 1204}, 004 (2012).

\bibitem{ihixs}
  C.~Anastasiou, S.~B\"uhler, F.~Herzog and A.~Lazopoulos,
  JHEP {\bf 1112}, 058 (2011).



\bibitem{Khachatryan:2014jba} 
  V.~Khachatryan {\it et al.}  [CMS Collaboration],
  arXiv:1412.8662 [hep-ex].
\bibitem{Aad:2014eha} 
  G.~Aad {\it et al.}  [ATLAS Collaboration],
  Phys.\ Rev.\ D {\bf 90}, no. 11, 112015 (2014)
  [arXiv:1408.7084 [hep-ex]].




\bibitem{formfactor}
 P.A.~Baikov, K.G.~Chetyrkin, A.V.~Smirnov, V.A.~Smirnov, M.~Steinhauser,
  Phys.\ Rev.\ Lett.\  {\bf 102}, 212002 (2009);
T.~Gehrmann, E.~W.~N.~Glover, T.~Huber, N.~Ikizlerli, C.~Studerus,
  JHEP {\bf 1006}, 094 (2010).
  
  
\bibitem{Anastasiou:2013mca}
  C.~Anastasiou, C.~Duhr, F.~Dulat, F.~Herzog and B.~Mistlberger,
  JHEP {\bf 1312}, 088 (2013)
  [arXiv:1311.1425 [hep-ph]];
\bibitem{Kilgore:2013gba}
  W.~B.~Kilgore,
  Phys.\ Rev.\ D {\bf 89} (2014) 7,  073008
  [arXiv:1312.1296 [hep-ph]].
  
\bibitem{Gehrmann:2011aa}
  T.~Gehrmann, M.~Jaquier, E.~W.~N.~Glover and A.~Koukoutsakis,
  JHEP {\bf 1202}, 056 (2012), 
 [arXiv:1112.3554 [hep-ph]].
  
\bibitem{Duhr:2013msa}
  C.~Duhr and T.~Gehrmann,
  Phys.\ Lett.\ B {\bf 727}, 452 (2013) 
  [arXiv:1309.4393 [hep-ph]];
  Y.~Li and H.~X.~Zhu,
  JHEP {\bf 1311}, 080 (2013)
  [arXiv:1309.4391 [hep-ph]].
  
\bibitem{Duhr:2014nda}
  C.~Duhr, T.~Gehrmann and M.~Jaquier,
  JHEP {\bf 1502} (2015) 077
  [arXiv:1411.3587 [hep-ph]];
  F.~Dulat and B.~Mistlberger,
  [arXiv:1411.3586 [hep-ph]].
  
  \bibitem{UV}
  O.~V.~Tarasov, A.~A.~Vladimirov and A.~Y.~.Zharkov,
  Phys.\ Lett.\ B {\bf 93}, 429 (1980);
  S.~A.~Larin and J.~A.~M.~Vermaseren,
  Phys.\ Lett.\ B {\bf 303}, 334 (1993)
  [hep-ph/9302208];
  T.~van Ritbergen, J.~A.~M.~Vermaseren and S.~A.~Larin,
  Phys.\ Lett.\ B {\bf 400}, 379 (1997)
  [hep-ph/9701390];
  M.~Czakon,
  Nucl.\ Phys.\ B {\bf 710}, 485 (2005)
  [hep-ph/0411261].
  \bibitem{NNLOXsec}
   C.~Anastasiou, S.~B\"uhler, C.~Duhr and F.~Herzog,
  JHEP {\bf 1211}, 062 (2012);
   M.~H\"oschele, J.~Hoff, A.~Pak, M.~Steinhauser, T.~Ueda,
  Phys.\ Lett.\ B {\bf 721}, 244 (2013);
  S.~B\"uhler and A.~Lazopoulos,
  JHEP {\bf 1310}, 096 (2013).



\bibitem{IR}
  S.~Moch, J.~A.~M.~Vermaseren and A.~Vogt,
  Nucl.\ Phys.\ B {\bf 688}, 101 (2004),
  [hep-ph/0403192];
  Nucl.\ Phys.\ B {\bf 691}, 129 (2004)
  [hep-ph/0404111].
  
  


 
  
    \bibitem{triplereal}
  C.~Anastasiou, C.~Duhr, F.~Dulat, B.~Mistlberger,
  JHEP {\bf 1307}, 003 (2013)
  [arXiv:1302.4379 [hep-ph]].
  
\bibitem{Zhu:2014fma}
  H.~X.~Zhu,
  JHEP {\bf 1502} (2015) 155
  [arXiv:1501.00236 [hep-ph]].
  
\bibitem{Anastasiou:2014vaa}
  C.~Anastasiou, C.~Duhr, F.~Dulat, E.~Furlan, T.~Gehrmann, F.~Herzog and B.~Mistlberger,
  Phys.\ Lett.\ B {\bf 737} (2014) 325
  [arXiv:1403.4616 [hep-ph]].
  
\bibitem{Li:2014bfa}
  Y.~Li, A.~von Manteuffel, R.~M.~Schabinger and H.~X.~Zhu,
  Phys.\ Rev.\ D {\bf 90} (2014) 5,  053006
  [arXiv:1404.5839 [hep-ph]].
  
\bibitem{Li:2014afw}
  Y.~Li, A.~von Manteuffel, R.~M.~Schabinger and H.~X.~Zhu,
  Phys.\ Rev.\ D {\bf 91} (2015) 3,  036008
  [arXiv:1412.2771 [hep-ph]].
  
\bibitem{Anastasiou:2014lda}
  C.~Anastasiou, C.~Duhr, F.~Dulat, E.~Furlan, T.~Gehrmann, F.~Herzog and B.~Mistlberger,
  [arXiv:1411.3584 [hep-ph]].
  
\bibitem{Moch:2005ky}
  S.~Moch and A.~Vogt,
  Phys.\ Lett.\ B {\bf 631}, 48 (2005)
  [hep-ph/0508265];
  E.~Laenen and L.~Magnea,
  Phys.\ Lett.\ B {\bf 632} (2006) 270
  [hep-ph/0508284];
  N.~A.~Lo Presti, A.~A.~Almasy and A.~Vogt,
  Phys.\ Lett.\ B {\bf 737} (2014) 120
  [arXiv:1407.1553 [hep-ph]];
  D.~de Florian, J.~Mazzitelli, S.~Moch and A.~Vogt,
  JHEP {\bf 1410} (2014) 176
  [arXiv:1408.6277 [hep-ph]].
  
\bibitem{otherSV}
  T.~Ahmed, M.~Mahakhud, N.~Rana and V.~Ravindran,
  Phys.\ Rev.\ Lett.\  {\bf 113} (2014) 11,  112002
  [arXiv:1404.0366 [hep-ph]];
  T.~Ahmed, M.~K.~Mandal, N.~Rana and V.~Ravindran,
  Phys.\ Rev.\ Lett.\  {\bf 113} (2014) 212003
  [arXiv:1404.6504 [hep-ph]];
  T.~Ahmed, N.~Rana and V.~Ravindran,
  JHEP {\bf 1410} (2014) 139
  [arXiv:1408.0787 [hep-ph]];
  T.~Ahmed, M.~K.~Mandal, N.~Rana and V.~Ravindran,
  JHEP {\bf 1502} (2015) 131
  [arXiv:1411.5301 [hep-ph]];
  M.~C.~Kumar, M.~K.~Mandal and V.~Ravindran,
  [arXiv:1412.3357 [hep-ph]];
  S.~Catani, L.~Cieri, D.~de Florian, G.~Ferrera and M.~Grazzini,
  Nucl.\ Phys.\ B {\bf 888} (2014) 75
  [arXiv:1405.4827 [hep-ph]].
  
\bibitem{nnlobabis} 
  C.~Anastasiou and K.~Melnikov,
  Nucl.\ Phys.\ B {\bf 646}, 220 (2002)
  [hep-ph/0207004].
  
\bibitem{reverse-unitarity}
  C.~Anastasiou and K.~Melnikov,
  Phys.\ Rev.\ D {\bf 67} (2003) 037501
  [hep-ph/0208115];
  C.~Anastasiou, L.~J.~Dixon and K.~Melnikov,
  Nucl.\ Phys.\ Proc.\ Suppl.\  {\bf 116} (2003) 193
  [hep-ph/0211141];
  C.~Anastasiou, L.~J.~Dixon, K.~Melnikov and F.~Petriello,
  Phys.\ Rev.\ Lett.\  {\bf 91} (2003) 182002
  [hep-ph/0306192];
  C.~Anastasiou, L.~J.~Dixon, K.~Melnikov and F.~Petriello,
  Phys.\ Rev.\ D {\bf 69} (2004) 094008
  [hep-ph/0312266].
  
  
  \bibitem{DiffEqs}
   A.~V.~Kotikov,
  Phys.\ Lett.\ B {\bf 259}, 314 (1991);
  Phys.\ Lett.\ B {\bf 267}, 123 (1991).
  
  

\bibitem{Wilson}
  K.~G.~Chetyrkin, B.~A.~Kniehl and M.~Steinhauser,
  Nucl.\ Phys.\ B {\bf 510}, 61 (1998);
  Y.~Schroder and M.~Steinhauser,
  JHEP {\bf 0601}, 051 (2006);
  K.~G.~Chetyrkin, J.~H.~Kuhn and C.~Sturm,
  Nucl.\ Phys.\ B {\bf 744}, 121 (2006).
  
  


\bibitem{nnlosoft} 
  S.~Catani, D.~de Florian and M.~Grazzini,
  JHEP {\bf 0105}, 025 (2001); 
  R.~V.~Harlander and W.~B.~Kilgore,
  Phys.\ Rev.\ D {\bf 64}, 013015 (2001)

\bibitem{mstw} 
  A.~D.~Martin, W.~J.~Stirling, R.~S.~Thorne and G.~Watt,
  Eur.\ Phys.\ J.\ C {\bf 63}, 189 (2009)
  
\bibitem{Hautmann:2002tu}
  F.~Hautmann,
  Phys.\ Lett.\ B {\bf 535} (2002) 159
  [hep-ph/0203140].



\bibitem{nlo}
  D.~Graudenz, M.~Spira and P.~M.~Zerwas,
  Phys.\ Rev.\ Lett.\  {\bf 70} (1993) 1372;
  S.~Dawson,
  Nucl.\ Phys.\ B {\bf 359} (1991) 283;
  A.~Djouadi, M.~Spira and P.~M.~Zerwas,
  Phys.\ Lett.\ B {\bf 264} (1991) 440;
  M.~Spira, A.~Djouadi, D.~Graudenz and P.~M.~Zerwas,
  Nucl.\ Phys.\ B {\bf 453} (1995) 17
  [hep-ph/9504378];
  R.~Harlander and P.~Kant,
  JHEP {\bf 0512}, 015 (2005)
  [hep-ph/0509189].
  C.~Anastasiou, S.~Beerli, S.~Bucherer, A.~Daleo and Z.~Kunszt,
  JHEP {\bf 0701}, 082 (2007)
  [hep-ph/0611236].
  U.~Aglietti, R.~Bonciani, G.~Degrassi and A.~Vicini,
  JHEP {\bf 0701}, 021 (2007)
  [hep-ph/0611266].
  R.~Bonciani, G.~Degrassi and A.~Vicini,
  JHEP {\bf 0711}, 095 (2007)
  [arXiv:0709.4227 [hep-ph]].
  C.~Anastasiou, S.~Bucherer and Z.~Kunszt,
  JHEP {\bf 0910}, 068 (2009)
  [arXiv:0907.2362 [hep-ph]].



  
  \bibitem{nnlo}
  R.~V.~Harlander and W.~B.~Kilgore,
  Phys.\ Rev.\ Lett.\  {\bf 88}, 201801 (2002)
  [hep-ph/0201206].
  V.~Ravindran, J.~Smith and W.~L.~van Neerven,
  Nucl.\ Phys.\ B {\bf 665}, 325 (2003)
  [hep-ph/0302135].

\bibitem{ewk}
  S.~Actis, G.~Passarino, C.~Sturm and S.~Uccirati,
  Nucl.\ Phys.\ B {\bf 811}, 182 (2009)
  [arXiv:0809.3667 [hep-ph]].
  S.~Actis, G.~Passarino, C.~Sturm and S.~Uccirati,
  Phys.\ Lett.\ B {\bf 670}, 12 (2008)
  [arXiv:0809.1301 [hep-ph]].
  U.~Aglietti, R.~Bonciani, G.~Degrassi and A.~Vicini,
  Phys.\ Lett.\ B {\bf 595}, 432 (2004)
  [hep-ph/0404071].
\bibitem{ewk3loop} 
  C.~Anastasiou, R.~Boughezal and F.~Petriello,
  JHEP {\bf 0904}, 003 (2009)
  [arXiv:0811.3458 [hep-ph]].

\bibitem{nnlotop}
  R.~V.~Harlander and K.~J.~Ozeren,
  JHEP {\bf 0911}, 088 (2009)
  [arXiv:0909.3420 [hep-ph]].
  A.~Pak, M.~Rogal and M.~Steinhauser,
  JHEP {\bf 1002}, 025 (2010)
  [arXiv:0911.4662 [hep-ph]].


\end{thebibliography}
\end{document}